%% file: conference_101719.tex
\documentclass[conference]{IEEEtran}
\IEEEoverridecommandlockouts
\usepackage{cite}
\usepackage{amsmath,amssymb}
\usepackage{algorithmic}
\usepackage{graphicx}
\usepackage{textcomp}
\usepackage{xcolor}
\usepackage{comment}
\usepackage{hyperref}

\usepackage{tikz}
\usetikzlibrary{shapes, positioning, arrows.meta}
\usetikzlibrary{fit}
\usepackage{array}
\usetikzlibrary{calc}

\usepackage{booktabs}
\usepackage{array}
\usepackage{makecell}

\usepackage{booktabs} 
\usepackage{multirow}  
\usepackage{tcolorbox}
\usepackage{pgfplots}
\pgfplotsset{compat=1.18}
\usepackage{pgf-pie}

\usepackage{pifont}
\newcommand{\cmark}{\ding{51}}%
\newcommand{\xmark}{\ding{55}}%

\definecolor{countrycolor}{RGB}{30, 100, 200} 
\newcommand{\sampleopacity}[1]{
    \ifnum #1 > 100 \def\op{0.9}
    \else\ifnum #1 > 50 \def\op{0.7}
    \else\ifnum #1 > 20 \def\op{0.5}
    \else\ifnum #1 > 10 \def\op{0.3}
    \else\ifnum #1 > 0 \def\op{0.15}
    \else \def\op{0} 
    \fi\fi\fi\fi\fi
}

\input{countries/World}
\usepackage{url}

\usetikzlibrary{shadows}
\usepackage{textcomp}

\usepackage{mathptmx}  
\usepackage{helvet}    
\usepackage{courier}   

\def\BibTeX{{\rm B\kern-.05em{\sc i\kern-.025em b}\kern-.08em
    T\kern-.1667em\lower.7ex\hbox{E}\kern-.125emX}}

\definecolor{KA1}{RGB}{140,86,75}  
\definecolor{KA2}{RGB}{44,160,44}   
\definecolor{KA3}{RGB}{214,39,40}   
\definecolor{KA4}{RGB}{148,103,189} 
\definecolor{KA5}{RGB}{31,119,180}  
\definecolor{KA6}{RGB}{188,189,34} 
\definecolor{KA7}{RGB}{127,127,127} 
\definecolor{KA8}{RGB}{227,119,194} 
\definecolor{KA9}{RGB}{255,127,14}  

\usepackage{pgf-pie}  
\newcommand{\piechart}[3][0.5]{
    \begin{tikzpicture}[baseline=-0.5ex]
        \def\radius{#1}
        \pgfmathsetmacro{\total}{0}
        
        \foreach \value/\color in {#2} {
            \pgfmathparse{\total + \value}
            \global\let\total\pgfmathresult
        }
        
        \pgfmathsetmacro{\startangle}{0}
        
        \foreach \value/\color in {#2} {
            \pgfmathsetmacro{\endangle}{\startangle + \value/\total*360}
            \draw[fill=\color, draw=black] 
                (0,0) -- (\startangle:\radius) 
                arc (\startangle:\endangle:\radius) -- cycle;
            \global\let\startangle\endangle
        }
        
    \end{tikzpicture}
}

\begin{document}

\title{CyberBridge: Bridging the Gap Between Cybersecurity Education and Industry
\thanks{Funding for this work was made available by the Swedish Research Council (VR) under grant 2021-05621, and the Swedish Knowledge Foundation (KKS).
All computations were executed on a Tesla T4 GPU, provided by the National Academic Infrastructure for Supercomputing in Sweden (NAISS), partially funded by the Swedish Research Council through grant agreement no. 2022-06725.}
}

\author{
\begin{tabular}{cc}
\begin{tabular}[t]{c}
\textbf{Arthur Nijdam}\\
\textit{Dept. of Electrical \& Information Technology}\\
\textit{Lund University}\\
Lund, Sweden\\
arthur.nijdam@eit.lth.se
\end{tabular}
&
\begin{tabular}[t]{c}
\textbf{Paul Stankovski Wagner}\\
\textit{Dept. of Electrical \& Information Technology}\\
\textit{Lund University}\\
Lund, Sweden\\
paul.stankovski\_wagner@eit.lth.se \vspace{0.2cm}
\end{tabular}
\\[1.5em]
\multicolumn{2}{c}{
\begin{tabular}[t]{c}
\textbf{Sara Ramezanian}\\
\textit{Dept. of Mathematics \& Computer Science}\\
\textit{Karlstad University}\\
Karlstad, Sweden\\
sara.ramezanian@kau.se
\end{tabular}
}
\end{tabular}
}



\maketitle

\begin{abstract}
This full research paper presents CyberBridge, a framework that automatically maps cybersecurity job descriptions to professional role profiles. As the cybersecurity landscape evolves rapidly, it is difficult for academic programs to align curricula with the competencies expected in practice. 

CyberBridge addresses this gap by decomposing a given vacancy description into its constituent Knowledge, Skill, and Task (KST) statements, embedding them using a sentence-BERT model, and matching them to the most semantically similar workforce profiles. A key contribution of our approach is its \emph{interpretability}. Rather than functioning as a black box, CyberBridge enables users to trace recommendations back to the specific competencies driving each match, providing a human-readable justification for the resulting mappings.


CyberBridge supports three primary use cases; (1) job recommendation, providing students with vacancies and closely matching professional roles based on their completed curriculum, (2) market analysis, enabling educators and curriculum developers to analyze which roles are currently in demand, and (3) curriculum planning, assessing which course program best prepares students for in-demand cybersecurity roles. 

As such, CyberBridge can be used by educational institutions as a practical tool for career guidance and evidence-based curriculum development in cybersecurity education. 
\end{abstract}

\begin{IEEEkeywords}
Career Development, Cybersecurity, LLM, NLP, Personalized Education, Recommender Systems
\end{IEEEkeywords}


\input{sections/introduction}

\input{sections/related_work}
\input{sections/methods}

\input{sections/results}

\input{sections/discussion}

\input{sections/conclusion}



\end{document}

%% file: sections/introduction.tex
\section{Introduction} \label{sec:introduction}
Cybersecurity is a rapidly evolving field, driven by the emergence of new threats such as Artificial Intelligence (AI)-based attacks \cite{guembe2022emerging}, Internet of Things vulnerabilities \cite{dave2023new}, and ransomware attacks \cite{teichmann2023evolution}.
At the same time, a persistent skills gap remains: in 2025, approximately two-thirds of cybersecurity teams reported lacking essential competencies \cite{ISC2CybersecurityWorkforce2025}. This highlights the need for stronger alignment between the regional labor market and cybersecurity education, where future cybersecurity professionals begin developing their skill set \cite{aldaajeh2022role,blavzivc2022changing, catal2023analysis}. 
However, curriculum developers and educators struggle to keep up with the fast-paced changes in the cybersecurity landscape \cite{  ramezanian2024cybersecurity}. 
Additionally, while workforce frameworks such as the National Initiative for Cybersecurity Education (NICE) published by the National Institute of Standards and Technology (NIST) \cite{NICE2025} provide structured descriptions of cybersecurity competencies and roles, academic curricula are often difficult to relate directly to these frameworks, limiting their practical use for students and curriculum designers.

To address these issues, we present CyberBridge, an end-to-end framework that connects cybersecurity curricula, standardized workforce frameworks, and real job advertisements (see Figure~\ref{fig:curriculum_flow}). CyberBridge supports three primary use cases: 
\begin{enumerate}
    \item \textbf{Job Recommendation:} help students identify which professional cybersecurity roles best match their completed coursework. 

    \item \textbf{Market Analysis:} enabling educators and program designers to analyze which cybersecurity roles are currently in demand in their region and to project how these demands evolve over time. 
    
    \item \textbf{Curriculum planning:} analyze how well existing and future curricula map onto in-demand cybersecurity roles. 
\end{enumerate}

\input{sections/figures/flowchartv2}

CyberBridge is an ontology-based Job Recommendation System (JRS) that maps unstructured job advertisements to the professional work roles defined in the NICE framework \cite{NICE2025}. The system follows a two-stage approach. First, an off-the-shelf Large Language Model (LLM) is used to extract structured Knowledge, Skill, and Task (KST) descriptions from the free-text job advertisement. These serve as a standardized intermediate representation, devoid of redundancy that might be present in the original phrasing of the advertisement, and more closely aligned with the structure of the NICE framework.
Subsequently, CyberBridge leverages the sentence-Bidirectional Encoder Representations from Transformers (sBERT) \cite{reimers2019sentence} model to derive semantic representations from the extracted KST, as well as of the KST associated with each NICE work role. By computing the similarity between these representations, the system establishes a bidirectional mapping between vacancies and NICE work roles: a vacancy can be linked to its nearest work role(s), and conversely, a given NICE work role can be matched to the most relevant available vacancies in the database. This symmetry enables both work role classification and downstream job recommendation, respectively, depending on whether the starting point is a job advertisement or a target role.

Therefore, we explicitly utilize the KST \textit{ontology} of the NICE framework, rather than relying on keyword matching or directly prompting a proprietary LLM. This ontology-centered design makes the job recommendation process more \textit{interpretable} and \textit{explainable}, since CyberBridge gives access to the specific Knowledge, Skill, or Task descriptions that contributed most strongly to the ranking of a given vacancy, e.g., whether this primarily driven by `hard skills' like C++ programming or by `soft skills' such as team work. This transparency is valuable in an educational context, as students and their guidance counselors get insight into which competencies are most important for a given cybersecurity role.
Additionally, the explainability of CyberBridge is beneficial to the workforce. It enables recruiters and hiring managers to assess how their vacancy is positioned within the NICE framework by exposing which workforce categories, e.g. Protect \& Defend versus Implement \& Operate, are most strongly associated with a given job posting. This insight, in turn, informs the design of vacancies targeted to the specific type of cybersecurity professional the organization aims to hire.
In summary, our methodology provides a transparent \textit{bridge} between job advertisements and standardized workforce frameworks, rather than relying on black-box machine learning.

Lastly, we demonstrate how CyberBridge can be coupled with the recently released CurricuLLM framework \cite{nijdam2026CLLM} to connect vacancies to academic curricula, as depicted in Figure \ref{fig:curriculum_flow}. This integration facilitates curriculum design based on current, regional labor market demand, enabling the use of CyberBridge as a \textit{curriculum-informed} JRS. 

In this work, we address the following research questions: 

\begin{itemize}
    \item RQ1: How can an interpretable matching be achieved between cybersecurity job advertisements and the professional roles defined in the NICE framework?

    \item RQ2: Which cybersecurity job roles are most in demand, and how does the demand differ geographically? 

    \item RQ3: How can labor market analyses performed by CyberBridge inform curriculum planning? 

\end{itemize}

The rest of this paper is organized as follows. Section \ref{sec:rel_work} gives an overview of related work. Section \ref{sec:methods} presents our proposed methodology, addressing RQ1. Section \ref{sec:results} presents both a quantitative performance evaluation of CyberBridge and experimental results from its application to market analysis (RQ2) and curriculum planning (RQ3). Section \ref{sec:discussion} discusses limitations and possible extensions of our tool. Finally, Section \ref{sec:conclusion} concludes our work.

%% file: sections/figures/flowchartv2.tex
\begin{figure}[h]
\centering
\resizebox{.35\textwidth}{!}{
\begin{tikzpicture}[
    node distance=2.5cm,
    arrow/.style={thick, -{Triangle[width=4mm,length=4mm]}, {Triangle[width=4mm,length=4mm]}-},
    thickarrow/.style={line width=4pt, -{Triangle[width=8mm,length=8mm]}, {Triangle[width=8mm,length=8mm]}-},
    curvedarrow/.style={thick, -{Triangle[width=4mm,length=4mm]}, bend angle=40},
    box/.style={draw, rectangle, minimum width=3cm, minimum height=1.2cm, align=center, font=\sffamily},
    widebox/.style={draw, rectangle, minimum width=6cm, minimum height=1.5cm, align=center, font=\sffamily}
]

\node[box] (job) {Job Advertisement};
\node[box, below=1.2cm of job] (ecsf) {Workforce Framework};
\node[box, below=1.2cm of ecsf] (academic) {Academic Curriculum};

\begin{scope}[line width=4pt]
    \draw[<->] ($(job.south)+(0,-0.1)$) -- node[midway, fill=white, text width=3cm, align=center, font=\bfseries\sffamily] {CyberBridge} ($(ecsf.north)+(0,0.1)$);
    \draw[<->] ($(ecsf.south)+(0,-0.1)$) -- node[midway, fill=white, text width=3cm, align=center, font=\bfseries\sffamily] {CurricuLLM} ($(academic.north)+(0,0.1)$);
\end{scope}


\draw[curvedarrow, bend right=80, -Latex] (job.west) to node[midway, left, rotate=90, anchor=north, font=\sffamily, yshift=20pt] {curriculum planning} (academic.west);
\draw[curvedarrow, bend right=85, -Latex] (academic.east) to node[midway, right, rotate=90, anchor=south, font=\sffamily, yshift=-20pt] {job recommendation} (job.east);
\draw[curvedarrow, bend left=40, -Latex] (job.east) to node[midway, right, rotate=90, anchor=south, font=\sffamily, xshift=-30pt, yshift=-15pt, fill=white, inner sep=2pt] {market analysis} (ecsf.east);

\end{tikzpicture}}
\caption{Linking job advertisements, standardized workforce frameworks, and academic curricula through CyberBridge and CurricuLLM \cite{nijdam2026CLLM}. }
\label{fig:curriculum_flow}
\end{figure}
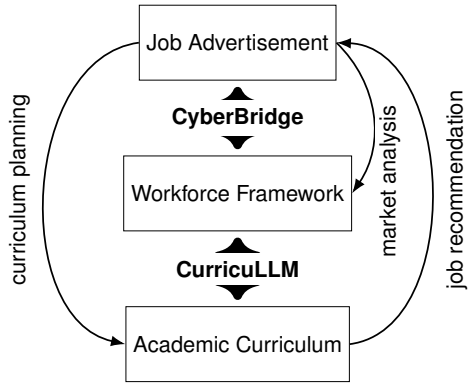

%% file: sections/related_work.tex
\section{Related Work} \label{sec:rel_work}

\subsection{Job Recommendation Systems} \label{sec:recc_systems}

Job Recommendation Systems aim to present users with vacancies that best match their profile and search criteria. To this end, JRS typically leverage information such as the content of the applicant's curriculum vitae, their past search behavior, or labor market data \cite{ccelik2025job}. 
In the STEM domain, professional competencies are often expressed using 
Knowledge, Skills, and Abilities or Task descriptions \cite{chang2019emerging,jia2018representation,kang2015job}. Because these KST descriptions define what a professional should know and do, they provide a natural foundation for JRS. 

This principle has been applied across computing disciplines. For example, Fern\'{a}ndez-Sanz et al. \cite{fernandezsanz2017eskills} developed a JRS for Computer Science that utilizes e-skills to align users with suitable roles. Within the cybersecurity domain, Dzurenda et al. \cite{dzurenda2024enhancing} proposed a JRS based on skills extracted from the European Cybersecurity Skills Framework (ECSF). 

Automating the analysis of vacancies for JRS is essential due to the time-consuming nature of manual analysis and the dynamic job market. Some JRS approaches use classical machine learning techniques such as Naive Bayes \cite{bachtiar2019employee} and K-means clustering \cite{bothmer2022investigating}. More recently, advances in Natural Language Processing (NLP), particularly transformer-based models such as BERT \cite{devlin2019bert}, have greatly improved the ability to process large text volumes. As a result, these models have become the new gold standard for JRS \cite{bothmer2022investigating,tamburri2020dataops,li2023joint}. 






\subsection{Cybersecurity Curriculum Development} \label{sec:curr_development}



Cybersecurity curricula require frequent updating due to the rapidly evolving threat landscape \cite{ramezanian2024cybersecurity}. Commonly used references for cybersecurity curriculum design are the Cybersecurity Curricula 2017 (CSEC2017) framework \cite{csec2017}, which organizes the domain into eight Knowledge Areas (KAs) to guide program design, and the Cyber Security Body of Knowledge (CyBOK) \cite{cybok}, which defines 21 Knowledge Areas. 

Aligning educational perspectives with workforce needs is a central challenge in curriculum development. Workforce frameworks, such as the NICE \cite{NICE2025} framework, describe cybersecurity professions in terms of roles and competencies.

Prior work has begun bridging educational and workforce perspectives. Ramezanian and Niemi \cite{ramezanian2024cybersecurity} mapped NICE Knowledge descriptions to CSEC2017, revealing gaps between academic coverage and workforce expectations. CurricuLLM \cite{nijdam2026CLLM} extends this idea by mapping course descriptions to NICE roles using LLMs. CyberBridge complements this approach by linking job advertisements to NICE, enabling labor market–aware curriculum analysis (Figure \ref{fig:curriculum_flow}).

%% file: sections/methods.tex
\section{Methods} \label{sec:methods}

\input{sections/figures/methodology_figv4}

\subsection{Datasets} \label{sec:data}

In this work, we use the following two datasets. 

\begin{itemize}
    \item \textbf{REWIRE \cite{ricci2022job}}: originally coined the Cybersecurity Job Ads Analyzer dataset, this dataset was developed for the REWIRE Horizon Europe project \cite{rewire}. It contains 936 European cybersecurity job descriptions, as well as one Afghan job description. 

    \item \textbf{LinkedIn \cite{LinkedIn}}: consists of approximately 1.3 million job listings scraped from LinkedIn in 2023 and 2024, made publicly available on \href{https://www.kaggle.com/datasets/asaniczka/1-3m-linkedin-jobs-and-skills-2024/discussion?sort=undefined}{Kaggle}. The dataset contains the job title, a link to the job description, the company, location of the job, date the listing was posted, the target city for the search, as well as a summary of the job listing and the required skills listed on LinkedIn. 
    6,020 samples in this dataset were deemed cybersecurity roles. 
\end{itemize}

We restricted our analysis to job descriptions and job titles, as these data fields were shared across both datasets. The REWIRE dataset contained approximately thirty ECSF-annotated samples, but these could not be utilized due to the lack of comparable annotations in the other dataset.
All job descriptions were normalized using a pre-processing function that converts text to lowercase, removes special characters, and cleans up extra spaces. 
All code developed for this work is publicly available in our Github repository \cite{github}.

\subsection{Extracting KST-statements} \label{sec:preprocessing}
\input{sections/figures/prompt}

We adopt an ontology-based method that connects the job advertisements contained in both datasets with the NICE job roles by breaking it down into Knowledge, Skill, and Task  descriptions. 
To this end, we use the procedure detailed in Figure \ref{fig:description_to_TKS}. The job description is first passed through an LLM with the prompt contained in Figure \ref{fig:kst-prompt} to extract KST statements. Each prompt variant (KNOWLEDGE, SKILL, TASK) uses the same base instructions (system prompt) with category-specific definitions appended to the system prompt (contained in \#Question). We employ DeepSeek-V3 as the LLM to do this extraction, since this LLM showed the best performance in \cite{nijdam2026CLLM}. 
As can be seen from the extracted KST statements in Figure \ref{fig:description_to_TKS}, all statements are taken from the original job description without alterations in their phrasing, except for the inclusion of `Knowledge of' and `Skill in', c.f. the phrasing of Knowledge and Skill descriptions in the NICE framework. For example, the sample that `Knowledge of ISO/IEC 27001' originates from appears under the heading \textit{Required Qualifications} in the job posting. 

Each resulting KST statement is embedded as a 768-dimensional vector using sBERT \cite{reimers2019sentence}, a sentence-level adaptation of BERT \cite{devlin2019bert}. Both models begin with tokenization, which splits text into smaller units such as words or subwords. BERT generates contextualized embeddings for each token in an input sequence, producing variable-length outputs that depend on the length of the input text. This complicates direct sentence-level similarity comparison because the embeddings do not share a fixed dimensional representation. sBERT addresses this limitation by applying pooling operations over the token embeddings to produce a single fixed-length sentence vector. sBERT embeddings are optimized for similarity comparison using cosine similarity, enabling efficient semantic comparison across statements of different lengths. 

This is particularly suitable for our setting: while we have access to elaborate, expert-defined, KST descriptions for all NICE work roles, we do not have labeled training data linking vacancies to roles. By embedding both the KST descriptions extracted from job advertisements and the official KST descriptions of NICE roles into the same semantic space, we can map a vacancy to its closest matching role(s) based on \textit{semantic similarity} rather than through supervised classification.

\subsection{Vacancy-NICE work role mapping with sBERT} \label{sec:classification}

\input{sections/figures/methodology_figv1}
Once the KST descriptions are embedded into the sBERT semantic space, each vacancy and each NICE work role can be represented in the same space by embedding of their respective Knowledge, Skill, and Task descriptions. This enables a structured comparison between vacancies and NICE work roles based on semantic similarity.
Figure \ref{fig:cyberbridge-nice-comprehensive} provides an intuitive geometric interpretation of the matching process in the embedding space. The formalization of similarity  underlying this visualization is as follows: 

Let a vacancy $v$ and a NICE work role $r$ consist of 3 sets corresponding to the sBERT embeddings of the K, S, and T descriptions, respectively, such that, $v=(K_v,S_v,T_v)$ and $r=(K_r,S_r,T_r)$, where for example $K_v = {\vec{k}_v^1,\vec{k}_v^2, ..., \vec{k}_v^8}$ if there are 8 Knowledge descriptions for vacancy $v$. 
We measure the semantic similarity between two embedding vectors using cosine similarity, defined as:
\begin{equation}\label{eq:cos_sim}
    sim(\vec{x},\vec{y}) = \frac{\vec{x}*\vec{y}}{||\vec{x}||*||\vec{y}||}
\end{equation}
This metric computes the cosine of the angle between vectors $\vec{x}$ and $\vec{y}$, providing a normalized measure of their semantic proximity. We adopt cosine similarity since sBERT is explicitly trained to produce embeddings that are comparable under this metric \cite{reimers2019sentence}. 

As shown in Figure \ref{fig:description_to_TKS}, some extracted KST descriptions are vague or not cybersecurity-specific, e.g., `Skill in establishing control-related processes and procedures'. To limit their influence, we apply a soft similarity threshold $\tau$, yielding the following \textit{thresholded} similarity function:
\begin{equation}\label{eq:soft_sim}
sim_{\tau}(\vec{x},\vec{y}) = \max\left(sim(\vec{x},\vec{y}) - \tau,  0\right)
\end{equation}
This ensures that only KST items from the vacancy that are close enough to NICE KST descriptions contribute to the final similarity score. Unless specified otherwise, we set $\tau=0.4$. 


Then, to compare $v$ and $r$, we compute: 
\begin{equation}\label{eq:role_sim}
    sim^{C}_{r\rightarrow v}(v,r) =  \frac{1}{|C_r|} \sum_{\vec{x}\in C_r} max_{\vec{y}\in C_v} sim_{\tau}(\vec{x},\vec{y})
\end{equation}
Where $C \in {K,S,T}$ denotes the competency category. For each vacancy-NICE work role pair, we compute the semantic similarity separately for Knowledge, Skill and Task embeddings. 
Every K, S, and T description in the vacancy is matched to the most semantically similar K, S, and T description contained in the work role using the cosine similarity between the corresponding sBERT embeddings. However, only samples exceeding the threshold $\tau$ contribute to the final score. The average of the best-match similarity scores quantifies how well the work role's competencies align with those of the vacancy.

Subsequently, the three category-level similarities are aggregated using predefined weights ($w^K, w^S$, and $w^T$): 
\begin{equation}\label{eq:weights}
    score(v,r) = w^K Sim^K_{r\rightarrow v}(v,r) + w^S Sim^S_{r\rightarrow v}(v,r) + w^TSim^T_{r\rightarrow v}(v,r)
\end{equation}
This weighted sum yields an overall similarity score between vacancy $v$ and NICE work role $r$. The role with the highest score is considered the best match for vacancy $v$. More generally, the function $score(v,r)$ induces a \textit{ranking} over all NICE work roles, reflecting their relative semantic proximity to the vacancy in the embedding space. This ranking-based interpretation enables downstream applications to consider the top-$k$ most relevant roles rather than a single role. Such flexibility is desirable because cybersecurity positions typically span multiple NICE work roles, reflecting the diversity of competencies expected of professionals in practice.

It is important to note that the labeled dataset used in this study is limited to 50 annotated samples. Since there are 41 distinct NICE work roles to differentiate, this equates to an average of just over one sample per role, rendering data-driven optimization of the weights $w^K$, $w^S$, and $w^T$ through $k$-fold cross-validation infeasible.  Therefore, the determination of the optimal weights is considered outside the scope of this paper. Unless specified otherwise, the results presented here are generated using equal weights, i.e., $w^K = w^S = w^T = \frac{1}{3}$.

\subsection{Quantitative Evaluation Methodology}

For quantitative evaluation, two human annotators labeled 25 vacancies with one or two NICE work roles, producing a dataset of 50 vacancies sampled equally from the REWIRE and LinkedIn datasets. CyberBridge’s performance is evaluated using three metrics commonly used in JRS research \cite{ccelik2025job}:
\begin{enumerate}
   
    \item \textbf{Mean Reciprocal Rank (MRR)}: MRR  reflects the position in the model’s ranking at which the annotator’s top label appears. For instance, an MRR of $1/2$ indicates that the role selected by the annotator appears, on average, in the second position of the model’s ranking. MRR can take on a value between 0 and 1, where higher is better. MRR is computed as:

\begin{equation}
\text{MRR} := \frac{1}{N} \sum_{n=1}^{N} \frac{1}{\text{r}_n},
\end{equation}

where $N$ is the total number of samples, 50 in our case, and $\text{r}_n$ is the rank of the annotator’s top label in the ranking generated by the model. 
 \item \textbf{Top-3 accuracy (Top3)}: Top-3 accuracy measures whether the annotator label appears among the three highest-ranked roles predicted by the model: 
\[
\mathrm{\text{Top3}} := \frac{1}{N} \sum_{n=1}^{N} 
\mathbb{I}(r_n \le 3),
\]
where $\mathbb{I}(\cdot)$ denotes the indicator function.

\item \textbf{Category Accuracy (Cat. Acc.)}: This metric evaluates whether the work role category of the annotator’s top label matches that of the model’s top prediction.

    
    
\end{enumerate}

%% file: sections/figures/methodology_figv4.tex
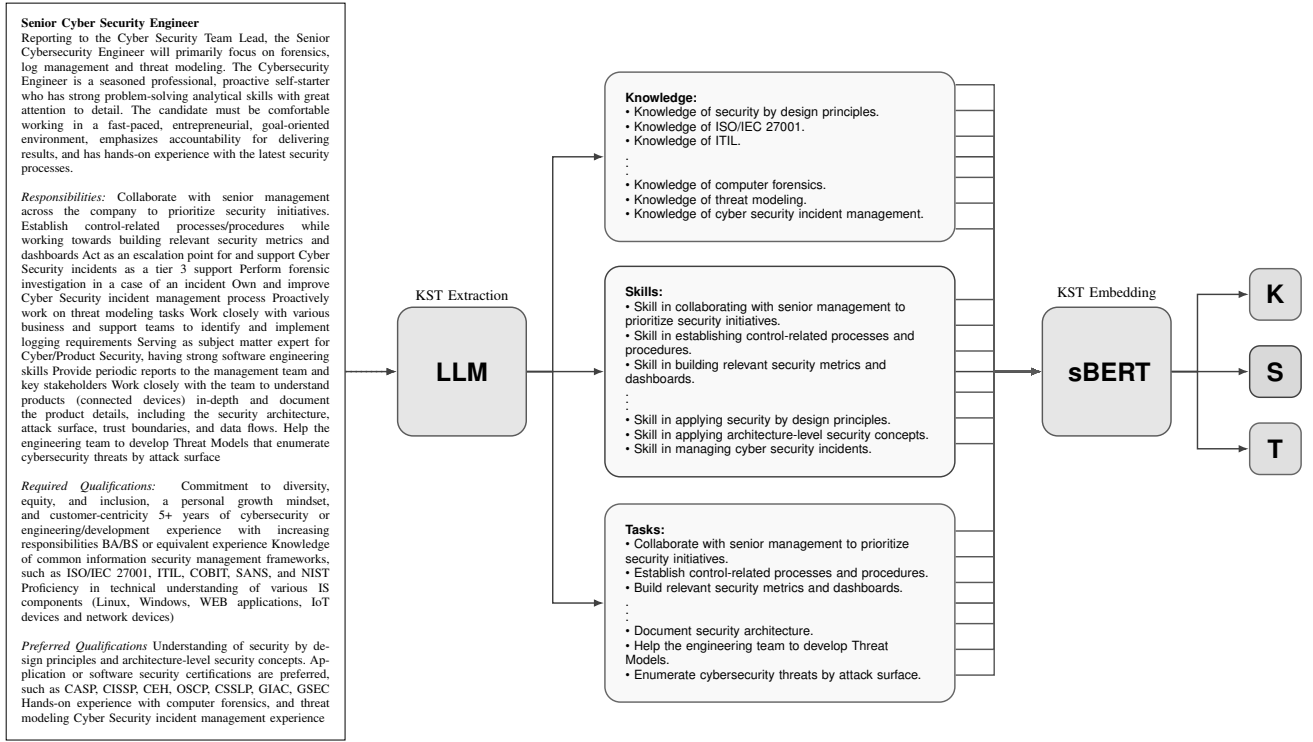
\begin{figure*}[htbp]
\centering
\resizebox{.95\textwidth}{!}{%
\begin{tikzpicture}[
    node distance=5mm and 10mm,
    box/.style={draw, rounded corners=3pt, minimum height=8mm, align=center},
    jobbox/.style={draw, rounded corners=3pt, minimum height=6mm, minimum width=5cm, align=center, font=\scriptsize},
    arrow/.style={-Latex, thick, shorten >=1pt, shorten <=1pt},
    descbox/.style={draw, text width=0.33\textwidth, align=justify, font=\scriptsize, inner sep=3mm},
    dashedbox/.style={draw, dashed, thick, rounded corners=5pt, minimum height=1.5cm, inner sep=5mm},
    legend/.style={font=\small, anchor=west},
    llmblock/.style={
        rectangle,
        draw=gray!70!black,
        thick,
        fill=gray!20,
        rounded corners=3mm,
        text width=7cm,
        align=center,
        minimum height=6.5cm,
        font=\bfseries,
        inner sep=3mm
    },
    labeltext/.style={
        font=\scriptsize,
        text=black
    },
    dot/.style={
        circle,
        fill=#1,
        inner sep=1.5pt,
        minimum size=3pt
    }
        titleblock/.style={
        rectangle,
        draw=blue!50!black,
        thick,
        fill=blue!20,
        rounded corners=3mm,
        text width=5cm,
        align=center,
        minimum height=1.5cm,
        font=\sffamily\bfseries
    },
    contentblock/.style={
        rectangle,
        draw=blue!50!black,
        thick,
        fill=white,
        rounded corners=3mm,
        text width=4cm,
        align=left,
        font=\sffamily\small,
        inner sep=4mm
    },
    llmblock/.style={
        rectangle,
        draw=gray!70!black,
        thick,
        fill=gray!20,
        rounded corners=3mm,
        text width=7cm,
        align=center,
        minimum height=2.5cm,
        font=\sffamily\Large\bfseries,
        inner sep=5mm
    },
    llmblock2/.style={
        rectangle,
        draw=gray!70!black,
        thick,
        fill=gray!20,
        rounded corners=3mm,
        text width=3cm,
        align=center,
        minimum height=2.5cm,
        font=\sffamily\Large\bfseries,
        inner sep=5mm
    },
    sbertblock/.style={
        rectangle,
        draw=gray!70!black,
        thick,
        fill=gray!20,
        rounded corners=3mm,
        text width=1.5cm,
        align=center,
        minimum height=2.5cm,
        font=\sffamily\Large\bfseries,
        inner sep=5mm
    },
    letterblock/.style={
        rectangle,
        draw=#1,
        thick,
        fill=#1!20,
        rounded corners=2mm,
        minimum width=1cm,
        minimum height=1cm,
        align=center,
        font=\sffamily\Large\bfseries,
        text=black
    },
    kblock/.style={
        rectangle,
        draw=gray!80!black,
        thick,
        fill=gray!5,
        rounded corners=3mm,
        text width=6cm,
        align=left,
        font=\sffamily\scriptsize,
        inner sep=4mm
    },
    sblock/.style={
        rectangle,
        draw=gray!50!black,
        thick,
        fill=gray!5,
        rounded corners=3mm,
        text width=6cm,
        align=left,
        font=\sffamily\scriptsize,
        inner sep=4mm
    },
    tblock/.style={
        rectangle,
        draw=gray!80!black,
        thick,
        fill=gray!5,
        rounded corners=3mm,
        text width=6cm,
        align=left,
        font=\sffamily\scriptsize,
        inner sep=4mm
    },
        weightcircle/.style={
        circle,
        draw=#1,
        thick,
        fill=#1!20,
        minimum size=1.2cm,
        align=center,
        font=\sffamily\bfseries,
        inner sep=2pt
    },
]


\node[descbox] (jobdesc) {
\textbf{Senior Cyber Security Engineer} \\
Reporting to the Cyber Security Team Lead, the Senior Cybersecurity Engineer will primarily focus on forensics, log management and threat modeling. The Cybersecurity Engineer is a seasoned professional, proactive self-starter who has strong problem-solving analytical skills with great attention to detail. The candidate must be comfortable working in a fast-paced, entrepreneurial, goal-oriented environment, emphasizes accountability for delivering results, and has hands-on experience with the latest security processes.\\ 

\textit{Responsibilities:}
Collaborate with senior management across the company to prioritize security initiatives.
Establish control-related processes/procedures while working towards building relevant security metrics and dashboards
Act as an escalation point for and support Cyber Security incidents as a tier 3 support
Perform forensic investigation in a case of an incident
Own and improve Cyber Security incident management process
Proactively work on threat modeling tasks
Work closely with various business and support teams to identify and implement logging requirements
Serving as subject matter expert for Cyber/Product Security, having strong software engineering skills
Provide periodic reports to the management team and key stakeholders
Work closely with the team to understand products (connected devices) in-depth and document the product details, including the security architecture, attack surface, trust boundaries, and data flows. Help the engineering team to develop Threat Models that enumerate cybersecurity threats by attack surface \\

\textit{Required Qualifications: }
Commitment to diversity, equity, and inclusion, a personal growth mindset, and customer-centricity
5+ years of cybersecurity or engineering/development experience with increasing responsibilities
BA/BS or equivalent experience
Knowledge of common information security management frameworks, such as ISO/IEC 27001, ITIL, COBIT, SANS, and NIST
Proficiency in technical understanding of various IS components (Linux, Windows, WEB applications, IoT devices and network devices) \\

\textit{Preferred Qualifications}
Understanding of security by design principles and architecture-level security concepts.
Application or software security certifications are preferred, such as CASP, CISSP, CEH, OSCP, CSSLP, GIAC, GSEC
Hands-on experience with computer forensics, and threat modeling
Cyber Security incident management experience

};

\coordinate (lowerstart) at ([yshift=0cm]jobdesc.south);
\node[sbertblock, above right=0cm and 1cm of jobdesc.east, anchor=west, label={[label distance=0mm]above:{\footnotesize KST Extraction}}] (llm) {LLM};

\draw[thick, dotted, gray!50!black, -{Latex[length=2mm, width=1.5mm]}] (jobdesc.east) -- ++(0.5cm,0) |- (llm.west);




\draw[thick, solid, gray!50!black, -{Latex[length=2mm, width=1.5mm]}] (jobdesc.east) -- ++(0.5cm,0) |- (llm.west);



\node[sblock, right=1.5cm of llm, align=left, text width=6cm] (scontent) {
    \textbf{Skills:}\\
    \textbullet\ Skill in collaborating with senior management to prioritize security initiatives.\\
    \textbullet\ Skill in establishing control-related processes and procedures.\\
    \textbullet\ Skill in building relevant security metrics and dashboards.\\
    \vdots
    \textbullet\ Skill in applying security by design principles.\\
    \textbullet\ Skill in applying architecture-level security concepts.\\
    \textbullet\ Skill in managing cyber security incidents.
};

\node[kblock, above right=1.25cm and 1.5cm of llm,  align=left, text width=6cm] (kcontent) {
    \textbf{Knowledge:}\\
    \textbullet\ Knowledge of security by design principles.\\
    \textbullet\ Knowledge of ISO/IEC 27001.\\
    \textbullet\ Knowledge of ITIL.\\
    \vdots
     \textbullet\ Knowledge of computer forensics.\\
    \textbullet\ Knowledge of threat modeling.\\
    \textbullet\ Knowledge of cyber security incident management.
};

\node[tblock,below=0.5cm of scontent, align=left, text width=6cm] (tcontent) {
    \textbf{Tasks:}\\
    \textbullet\ Collaborate with senior management to prioritize security initiatives.\\
    \textbullet\ Establish control-related processes and procedures.\\
    \textbullet\ Build relevant security metrics and dashboards.\\
    \vdots
    \textbullet\ Document security architecture.\\
    \textbullet\ Help the engineering team to develop Threat Models.\\
    \textbullet\ Enumerate cybersecurity threats by attack surface.
};

\node[sbertblock, right=10cm of llm, label={[label distance=0mm]above:{\footnotesize KST Embedding}}] (sbert) {sBERT};

\draw[thick, solid, gray!50!black, -{Latex[length=2mm, width=1.5mm]}] (llm.east) -- ++(0.5cm,0) |- (kcontent.west);
\draw[thick, solid, gray!50!black, -{Latex[length=2mm, width=1.5mm]}] (llm.east) -- ++(0.5cm,0) |- (scontent.west);
\draw[thick, solid, gray!50!black, -{Latex[length=2mm, width=1.5mm]}] (llm.east) -- ++(0.5cm,0) |- (tcontent.west);

\draw[thick, solid, gray!50!black, -{Latex[length=2mm, width=1.5mm]}] (kcontent.east) -- ++(0.75cm,0) |- (sbert.west);
\draw[thick, solid, gray!70!black, -{Latex[length=2mm, width=1.5mm]}] ([yshift=-0.4cm]kcontent.east) -- ++(0.75cm,0) |- (sbert.west);
\draw[thick, solid, gray!70!black, -{Latex[length=2mm, width=1.5mm]}] ([ yshift=-0.9cm]kcontent.east) -- ++(0.75cm,0) |- (sbert.west);
\draw[thick, solid, gray!70!black, -{Latex[length=2mm, width=1.5mm]}] ([yshift=0.4cm]kcontent.east) -- ++(0.75cm,0) |- (sbert.west);
\draw[thick, solid, gray!70!black, -{Latex[length=2mm, width=1.5mm]}] ([yshift=-1.4cm]kcontent.east) -- ++(0.75cm,0) |- (sbert.west);
\draw[thick, solid, gray!70!black, -{Latex[length=2mm, width=1.5mm]}] ([ yshift=1.4cm]kcontent.east) -- ++(0.75cm,0) |- (sbert.west);
\draw[thick, solid, gray!70!black, -{Latex[length=2mm, width=1.5mm]}] ([ yshift=0.9cm]kcontent.east) -- ++(0.75cm,0) |- (sbert.west);

\draw[thick, solid, gray!50!black, -{Latex[length=2mm, width=1.5mm]}] (scontent.east) -- ++(0.75cm,0) |- (sbert.west);
\draw[thick, solid, gray!50!black, -{Latex[length=2mm, width=1.5mm]}] (tcontent.east) -- ++(0.75cm,0) |- (sbert.west);

\draw[thick, solid, gray!70!black, -{Latex[length=2mm, width=1.5mm]}] ([yshift=1.4cm]scontent.east) -- ++(0.75cm,0) |- (sbert.west);
\draw[thick, solid, gray!70!black, -{Latex[length=2mm, width=1.5mm]}] ([yshift=0.9cm]scontent.east) -- ++(0.75cm,0) |- (sbert.west);
\draw[thick, solid, gray!70!black, -{Latex[length=2mm, width=1.5mm]}] ([yshift=0.4cm]scontent.east) -- ++(0.75cm,0) |- (sbert.west);
\draw[thick, solid, gray!70!black, -{Latex[length=2mm, width=1.5mm]}] ([yshift=-0.4cm]scontent.east) -- ++(0.75cm,0) |- (sbert.west);
\draw[thick, solid, gray!70!black, -{Latex[length=2mm, width=1.5mm]}] ([yshift=-0.9cm]scontent.east) -- ++(0.75cm,0) |- (sbert.west);
\draw[thick, solid, gray!70!black, -{Latex[length=2mm, width=1.5mm]}] ([yshift=-1.4cm]scontent.east) -- ++(0.75cm,0) |- (sbert.west);

\draw[thick, solid, gray!70!black, -{Latex[length=2mm, width=1.5mm]}] ([yshift=1.4cm]tcontent.east) -- ++(0.75cm,0) |- (sbert.west);
\draw[thick, solid, gray!70!black, -{Latex[length=2mm, width=1.5mm]}] ([yshift=0.9cm]tcontent.east) -- ++(0.75cm,0) |- (sbert.west);
\draw[thick, solid, gray!70!black, -{Latex[length=2mm, width=1.5mm]}] ([yshift=0.4cm]tcontent.east) -- ++(0.75cm,0) |- (sbert.west);
\draw[thick, solid, gray!70!black, -{Latex[length=2mm, width=1.5mm]}] ([yshift=-0.4cm]tcontent.east) -- ++(0.75cm,0) |- (sbert.west);
\draw[thick, solid, gray!70!black, -{Latex[length=2mm, width=1.5mm]}] ([yshift=-0.9cm]tcontent.east) -- ++(0.75cm,0) |- (sbert.west);
\draw[thick, solid, gray!70!black, -{Latex[length=2mm, width=1.5mm]}] ([yshift=-1.4cm]tcontent.east) -- ++(0.75cm,0) |- (sbert.west);


\node[letterblock=gray!80!black,, below right=1.5cm and 1.5cm of sbert.east, anchor=west] (tletter_right) {T};
\node[letterblock=gray!50!black, right= 1.5cm of sbert.east, anchor=west] (sletter_right) {S};
\node[letterblock=gray!80!black, above=1.5cm of sletter_right.west, anchor=west] (kletter_right) {K};

\draw[thick, solid, gray!50!black, -{Latex[length=2mm, width=1.5mm]}] (sbert.east) -- ++(0.5cm,0) |- (kletter_right.west);
\draw[thick, solid, gray!50!black, -{Latex[length=2mm, width=1.5mm]}] (sbert.east) -- ++(0.5cm,0) |- (sletter_right.west);
\draw[thick, solid, gray!50!black, -{Latex[length=2mm, width=1.5mm]}] (sbert.east) -- ++(0.5cm,0) |- (tletter_right.west);
\end{tikzpicture}
    }%
\caption{A job description from the REWIRE dataset is mapped to separate K, S and T embeddings using CyberBridge. First, the job description is fed through an LLM to extract Knowledge, Skill, and Task descriptions (consistent with the NICE framework), then, these descriptions along with the original vacancy are passed through the sBERT model to extract embeddings. The K, S, and T embeddings can now be compared to the NICE work role KST embeddings.} 
\label{fig:description_to_TKS}
\end{figure*}

%% file: sections/figures/prompt.tex
\begin{figure}[ht]
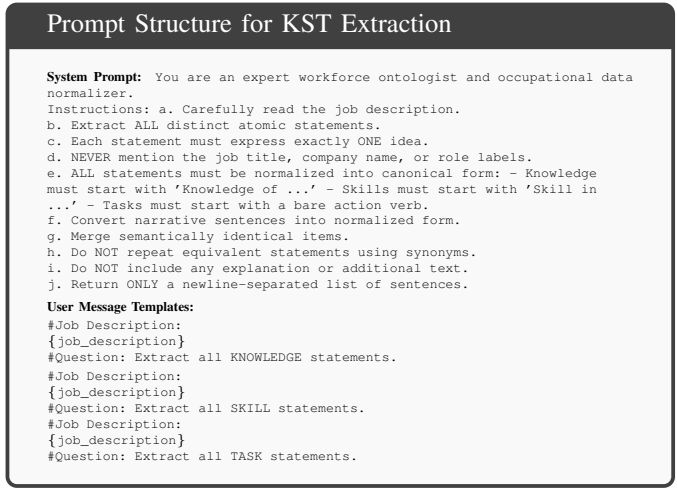

\centering
\begin{tcolorbox}[colback=gray!5!white,colframe=black!75!white,
  title=Prompt Structure for KST Extraction,
  fontupper=\tiny]
\textbf{System Prompt:}
\begin{ttfamily}
You are an expert workforce ontologist and occupational data normalizer.

Instructions:
a. Carefully read the job description.

b. Extract ALL distinct atomic statements.

c. Each statement must express exactly ONE idea.

d. NEVER mention the job title, company name, or role labels.

e. ALL statements must be normalized into canonical form:
   - Knowledge must start with 'Knowledge of ...'
   - Skills must start with 'Skill in ...'
   - Tasks must start with a bare action verb.
   
f. Convert narrative sentences into normalized form.

g. Merge semantically identical items.

h. Do NOT repeat equivalent statements using synonyms.

i. Do NOT include any explanation or additional text.

j. Return ONLY a newline-separated list of sentences.
\end{ttfamily}

\vspace{0.5em}
\textbf{User Message Templates:}

\vspace{0.2em}
\texttt{\#Job Description: \\ \{job\_description\} \\ \#Question: Extract all KNOWLEDGE statements.}

\vspace{0.2em}
\texttt{\#Job Description: \\ \{job\_description\} \\ \#Question: Extract all SKILL statements.}

\texttt{\#Job Description: \\ \{job\_description\} \\ \#Question: Extract all TASK statements.}

\end{tcolorbox}
\caption{Prompt used for DeepSeekV3 to extract KSTs from job descriptions, where \{job\_description\} is replaced with the actual job description.}
\label{fig:kst-prompt}
\end{figure}

%% file: sections/figures/methodology_figv1.tex
\begin{figure*}[htbp]
\centering
\resizebox{.95\textwidth}{!}{%
\begin{tikzpicture}[
    node distance=5mm and 10mm,
    box/.style={draw, rounded corners=3pt, minimum height=8mm, align=center},
    jobbox/.style={draw, rounded corners=3pt, minimum height=8mm, minimum width=2.9cm, align=center, font=\scriptsize},
    arrow/.style={-Latex, thick, shorten >=1pt, shorten <=1pt},
    descbox/.style={draw, text width=0.25\textwidth, align=justify, font=\small, inner sep=3mm},
    dashedbox/.style={draw, dashed, thick, rounded corners=5pt, minimum height=1.5cm, inner sep=5mm},
    legend/.style={font=\small, anchor=west},
    llmblock/.style={
        rectangle,
        draw=gray!70!black,
        thick,
        fill=gray!20,
        rounded corners=3mm,
        text width=7cm,
        align=center,
        minimum height=6.5cm,
        font=\bfseries,
        inner sep=3mm
    },
    labeltext/.style={
        font=\scriptsize,
        text=black
    },
    dot/.style={
        circle,
        fill=#1,
        inner sep=1.5pt,
        minimum size=3pt
    }
]


\node[descbox] (jobdesc) {
\textbf{Senior Cyber Security Engineer}\\
Reporting to the Cyber Security Team Lead, the Senior Cybersecurity Engineer will primarily focus on forensics, log management and threat modeling. The Cybersecurity Engineer is a seasoned professional, proactive self-starter who has strong problem-solving analytical skills with great attention to detail. The candidate must be comfortable working in a fast-paced, entrepreneurial, goal-oriented environment, emphasizes accountability for delivering results, and has hands-on experience with the latest security processes. 
};

\node[llmblock, right=20 mm of jobdesc, anchor=west, xshift=1cm] (bridge) {
    
    \begin{tikzpicture}[scale=0.6, transform shape]

   \tikzset{
        dot/.style={circle, fill=#1, draw=black!80!#1, thick, minimum size=4mm, inner sep=0pt},
        labeltext/.style={font=\tiny, anchor=west},
        cross/.style={draw, cross out, minimum size=3mm, thick, inner sep=0pt}
    }
    
        \fill[white] (-1, -1) rectangle (10.5, 10.5);
        \node[dot=blue!50] (j1) at (6.062404984882988,2.5537004328704094) {};
        \node[dot=blue!50] (j2) at (4.247918264299129,0.8008835470460827) {};
        \node[dot=blue!50] (j3) at (4.3352644645485165,0.2180983360017959) {};
        \node[dot=blue!50] (j4) at (2.647369233127428,10.0) {};
        \node[dot=blue!50] (j5) at (2.5717246137138225,9.590073501733169) {};
        \node[dot=blue!50] (j6) at (3.7298661787088823,1.4180866210784664) {};
        \node[dot=blue!50] (j7) at (3.802595165702768,0.7544781349568294) {};
        \node[dot=blue!50] (j8) at (4.860139545797399,1.3947793989822275) {};
        \node[dot=blue!50] (j9) at (9.462669566304033,0.765300180366473) {};
        \node[dot=blue!50] (j10) at (8.93393151090106,0.6010822216513685) {};
        \node[dot=blue!50] (j11) at (8.875851483786635,1.0989048656970162) {};
        \node[dot=blue!50] (j12) at (7.128752731895322,2.9783840843127707) {};
        \node[dot=blue!50] (j13) at (7.5058273316293915,2.6088044892224986) {};
        \node[dot=blue!50] (j14) at (6.506736914547303,2.63687610333589) {};
        \node[dot=blue!50] (j15) at (10.0,1.205558577126173) {};
        \node[dot=blue!50] (j16) at (9.41811978175404,0.0) {};
        
        \node[dot=green!50] (j17) at (6.521739009262597,4.844971822586555) {};
        \node[dot=green!50] (j18) at (6.9133940186438645,4.489342625119259) {};
        \node[dot=green!50] (j19) at (6.107139338123908,6.867071962709357) {};
        \node[dot=green!50] (j20) at (6.982934296063283,5.573323511200926) {};
        \node[dot=green!50] (j21) at (6.304144277336708,6.476690679027526) {};
        \node[dot=green!50] (j22) at (8.81644735836623,1.7053877696656718) {};
        \node[dot=green!50] (j23) at (7.678307852125686,5.687112831479858) {};
        \node[dot=green!50] (j24) at (5.727406450284674,5.759348394747517) {};
        \node[dot=green!50] (j25) at (7.445587118130255,1.1937499001893097) {};
        
        \node[dot=orange!50] (j26) at (3.608996667334938,7.857259442547598) {};
        \node[dot=orange!50] (j27) at (4.1358184066139065,7.996856385306224) {};
        \node[dot=orange!50] (j28) at (8.018053209364545,3.4021796145276912) {};
        \node[dot=orange!50] (j29) at (3.9622520279631006,5.168290999151781) {};
        \node[dot=orange!50] (j30) at (5.450999844469637,4.867278211520184) {};
        \node[dot=orange!50] (j31) at (6.306258388975145,5.540477649477631) {};
        \node[dot=orange!50] (j32) at (5.301962649338367,4.382428390677522) {};
        
        \node[dot=purple!50] (j33) at (2.2745641329002457,7.638363214182563) {};
        \node[dot=purple!50] (j34) at (0.4738838224148921,7.028896894530101) {};
        \node[dot=purple!50] (j35) at (1.5879402835327507,7.368520734843947) {};
        \node[dot=purple!50] (j36) at (3.3304590932669735,5.443623089281176) {};
        \node[dot=purple!50] (j37) at (2.2299368079235977,6.843596351423061) {};
        \node[dot=purple!50] (j38) at (2.7062375879128635,7.431894249489768) {};
        \node[dot=purple!50] (j39) at (2.888975650507991,5.695947936918731) {};
        
        \node[dot=red!50] (j40) at (1.0373290851668087,7.143133914235355) {};
        \node[dot=red!50] (j41) at (0.0,7.074906748297743) {};
        \node[cross, draw=black, thick] (cross) at (3.3, 8.65) {};
        \draw[dotted, thick, black!70] (cross.center) -- (j38.center);
        \draw[dotted, thick, black!70] (cross.center) -- (j5.center);
        \draw[dotted, thick, black!70] (cross.center) -- (j26.center);

    \end{tikzpicture}
};

\draw[arrow] (jobdesc.east) -- ++(1.7cm,0) |- (8.35, 2.35);

\node[fill=white, draw=gray!50, rounded corners=3pt, inner sep=4pt, font=\normalsize\bfseries] 
    at ([xshift=1.5cm, yshift=0cm]jobdesc.east) {CyberBridge};
\coordinate[above right=35mm and 35mm of bridge.east] (gridstart);

\node[jobbox, fill=blue!20] at ($(gridstart) + (0,0)$) (job1) {Communications Security\\(COMSEC) Management};
\node[jobbox, fill=blue!20, right=2mm of job1] (job2) {Cybersecurity\\Policy and Planning};
\node[jobbox, fill=blue!20, right=2mm of job2] (job3) {Cybersecurity\\Workforce Management};
\node[jobbox, fill=blue!20, right=2mm of job3] (job4) {Cybersecurity\\Curriculum Development};
\node[jobbox, fill=blue!20, right=2mm of job4] (job5) {Cybersecurity\\Instruction};

\node[jobbox, fill=blue!40, below=2mm of job1] (job6) {\textbf{Executive}\\ \textbf{Cybersecurity Leadership}};
\node[jobbox, fill=blue!20, right=2mm of job6] (job7) {Cybersecurity\\Legal Advice};
\node[jobbox, fill=blue!20, right=2mm of job7] (job8) {Privacy\\Compliance};
\node[jobbox, fill=blue!20, right=2mm of job8] (job9) {Product Support\\Management};
\node[jobbox, fill=blue!20, right=2mm of job9] (job10) {Program\\Management};

\node[jobbox, fill=blue!20, below=2mm of job6] (job11) {Secure Project\\Management};
\node[jobbox, fill=blue!20, right=2mm of job11] (job12) {Security Control\\Assessment};
\node[jobbox, fill=blue!20, right=2mm of job12] (job13) {Systems\\Authorization};
\node[jobbox, fill=blue!20, right=2mm of job13] (job14) {Systems Security\\Management};
\node[jobbox, fill=blue!20, right=2mm of job14] (job15) {Technology Portfolio\\Management};

\node[jobbox, fill=blue!20, below=2mm of job11] (job16) {Technology Program\\Auditing};
\node[jobbox, fill=green!20, right=2mm of job16] (job17) {Cybersecurity\\Architecture};
\node[jobbox, fill=green!20, right=2mm of job17] (job18) {Enterprise\\Architecture};
\node[jobbox, fill=green!20, right=2mm of job18] (job19) {Operational Technology\\Cybersecurity Engineering};
\node[jobbox, fill=green!20, right=2mm of job19] (job20) {Secure Software\\Development};

\node[jobbox, fill=green!20, below=2mm of job16] (job21) {Secure Systems\\Development};
\node[jobbox, fill=green!20, right=2mm of job21] (job22) {Software Security\\Assessment};
\node[jobbox, fill=green!20, right=2mm of job22] (job23) {Systems\\Requirements Planning};
\node[jobbox, fill=green!20, right=2mm of job23] (job24) {Systems Testing\\and Evaluation};
\node[jobbox, fill=green!20, right=2mm of job24] (job25) {Technology Research\\and Development};

\node[jobbox, fill=orange!20, below=2mm of job21] (job26) {Data Analysis};
\node[jobbox, fill=orange!20, right=2mm of job26] (job27) {Database\\Administration};
\node[jobbox, fill=orange!20, right=2mm of job27] (job28) {Knowledge\\Management};
\node[jobbox, fill=orange!20, right=2mm of job28] (job29) {Network\\Operations};
\node[jobbox, fill=orange!20, right=2mm of job29] (job30) {Systems\\Administration};

\node[jobbox, fill=orange!40, below=2mm of job26] (job31) {\textbf{Systems Security}\\ \textbf{Analysis}};
\node[jobbox, fill=orange!20, right=2mm of job31] (job32) {Technical\\Support};
\node[jobbox, fill=purple!20, right=2mm of job32] (job33) {Defensive\\Cybersecurity};
\node[jobbox, fill=purple!20, right=2mm of job33] (job34) {Digital\\Forensics};
\node[jobbox, fill=purple!40, below=2mm of job31] (job35) {\textbf{Incident}\\ \textbf{Response}};

\node[jobbox, fill=purple!20, right=2mm of job34] (job36) {Infrastructure\\Support};
\node[jobbox, fill=purple!20, right=2mm of job35] (job37) {Insider Threat\\Analysis};
\node[jobbox, fill=purple!20, right=2mm of job37] (job38) {Threat\\Analysis};
\node[jobbox, fill=purple!20, right=2mm of job38] (job39) {Vulnerability\\Analysis};
\node[jobbox, fill=red!20, right=2mm of job39] (job40) {Cybercrime\\Investigation};

\node[jobbox, fill=red!20, below=2mm of job35] (job41) {Digital Evidence\\Analysis};

\draw[arrow] (8.4, 2.9) -- ++(5.95cm,0) |- (job6.west);
\draw[arrow] (8.95, 1.9) -- ++(5.25cm,0) |- (job31.west);
\draw[arrow] (8.4, 1.6) -- ++(5.5cm,0) |- (job35.west);

\node[above=1mm of jobdesc.north, anchor=south, font=\bfseries] {Job Description};
\node[above=2mm of bridge.north, anchor=south, font=\bfseries] {CyberBridge Embedding Space};
\node[above right=1mm and 40mm of job1.north west, anchor=south west, font=\bfseries] {NICE Framework Work Roles };


\end{tikzpicture}
}
\caption{A birds-eye overview of CyberBridge. A job description is mapped onto the cyberbridge embedding space, where it can be linked to the closest NICE framework work roles. All NICE work roles are listed on the right-hand side, color-coded by category: \textcolor{blue!60}{$\blacksquare$} Oversight \& Governance, \textcolor{green!60}{$\blacksquare$} Design \& Development, \textcolor{orange!60}{$\blacksquare$} Implementation \& Operation, \textcolor{purple!60}{$\blacksquare$} Protection \& Defense, and \textcolor{red!60}{$\blacksquare$} Investigation. 
}
\label{fig:cyberbridge-nice-comprehensive}
\end{figure*}
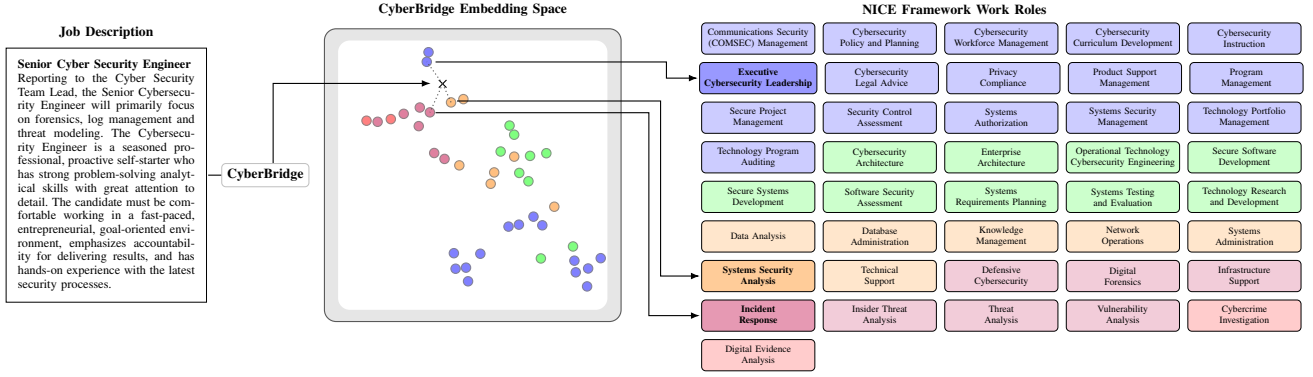

%% file: sections/results.tex
\section{Results} \label{sec:results}

\subsection{Performance Assessment}
\begin{table}[ht!]
\centering
\caption{CyberBridge ablations. Metrics were computed on human-annotated data.}
\label{tab:results_ablations}
\begin{tabular}{lcccccc}
\toprule
\textbf{Method} & \multicolumn{3}{c}{\textbf{Components}} & \multicolumn{3}{c}{\textbf{Metrics}} \\
\cmidrule(lr){2-4} \cmidrule(lr){5-7}
& \textbf{K} & \textbf{S} & \textbf{T} & \textbf{MRR} & \textbf{Top3} & \textbf{Cat. Acc.} \\

\midrule
\multicolumn{6}{l}{\textit{Ablations}} \\
\midrule
Description  & \xmark & \xmark & \xmark  & 0.08 & 6 & 4 \\
Ours K & \cmark & \xmark & \xmark & 0.17 & 16 & 36 \\
Ours S & \xmark & \cmark & \xmark & 0.19 & 14 & 30 \\
Ours T & \xmark & \xmark & \cmark & 0.28 & 28 & 44 \\
Ours (equal K/S/T) & \cmark & \cmark & \cmark & 0.27 & 28 & 44 \\

\bottomrule
\end{tabular}
\end{table}

Here, we evaluate the performance of CyberBridge by comparing different variants of our framework and several baselines with human annotator ground truth.

\begin{table}[ht!]
\centering
\caption{Zero-shot performance on human-annotated data for several baseline models.}
\label{tab:results_baselines}
\begin{tabular}{lccc}
\toprule
\textbf{Method} &  \multicolumn{3}{c}{\textbf{Metrics}} \\
\cmidrule(lr){2-4} 
 & \textbf{MRR} & \textbf{Top3}  & \textbf{Cat. Acc.} \\

\midrule

\multicolumn{4}{l}{\textit{Baselines}} \\
\midrule
Qwen2.5-7B-Instruct  & N/A & 2 & 20  \\
ChatGPT-4o-mini  & 0.14 & 7 & 24 \\
ChatGPT-5-mini  & 0.36 & 33 & 42 \\
DeepSeek-V3  & 0.39 & 37 &  50 \\
\bottomrule
\end{tabular}
\end{table}

We first analyze how the individual Knowledge, Skill, and Task components of CyberBridge influence its overall performance on human-labeled data. These component-wise evaluations, known as \textit{ablations}, are presented in Table \ref{tab:results_ablations}. Firstly, we use CyberBridge without extracting KST-descriptions, where we directly map the overall vacancy description to a NICE work role description. As indicated for `Description' in Table \ref{tab:results_ablations}, this approach performs poorly (Top3 = 6\%, MRR = 0.08). This is likely due to the verbosity and redundancy of job advertisements, compounded by the fact that sBERT was designed to process sentences rather than paragraphs of text.

Basing the prediction on Knowledge, Skills, or Tasks yields stronger results: The Knowledge-only and Skill-only versions achieve Top3 scores of 16\% and 14\%, respectively, while Task-based matching achieves 28\%. This makes sense, since task descriptions often provide the most concrete information about role requirements. Combining Knowledge, Skills, and Tasks with equal weighting yields a comparable performance. 

For reference, we also evaluate several state-of-the-art LLMs through \textit{zero-shot prompting}. In the zero-shot setting, the model receives the job description together with a list of all NICE work roles and their summaries as listed in the NICE framework \cite{NICE2025}. 
The model is then asked to rank the roles according to their relevance to the job description. 
It is important to note that it was not feasible to inform these zero-shot models of the full NICE framework, including all KST descriptions per work role, due to token length limits and the inability to attach external files in DeepSeek and Qwen's APIs. We recommend that future work explores this avenue for OpenAI models, as their Responses API provides functionality for file inclusion up to 512 MB. 

Among the zero-shot baselines, DeepSeek-V3 \cite{liu2024deepseek} achieves the highest performance, reaching 37\% Top3, 50\% category accuracy and an MRR of 0.39. CyberBridge outperforms two out of four zero-shot baselines (ChatGPT-4o-mini and Qwen2.5-7B-Instruct) while being built a computationally efficient sBERT architecture that requires substantially fewer resources.However, a performance gap remains relative to DeepSeek-V3 and ChatGPT-5-mini.  
This gap may be partially explained by the human annotation process itself. Annotators reported that it took considerable effort to map a vacancy to the most appropriate work role. Due to the difficulty of this task, it is plausible that annotators relied primarily on high-level role understanding rather than conducting a detailed analysis of the underlying Knowledge, Skills and Tasks. In this case, the higher correspondence between LLM predictions and human judgments does not imply that zero-shot prompting leads to a deeper semantic understanding of vacancy-to-role mapping. 
CyberBridge, by contrast, grounds its predictions in KST descriptions, which offers interpretability (see Section \ref{sec:interpretability}) at the cost of lower agreement with human labels.

\subsection{Interpretability}
\label{sec:interpretability}

To illustrate how CyberBridge results in interpretable mappings, we consider Figure \ref{fig:cyberbridge-nice-comprehensive}. This figure presents an illustrative two-dimensional projection of the high-dimensional sBERT embedding space, showing how NICE work roles can be related to a given vacancy using CyberBridge. The vacancy corresponds to the example previously analyzed in Figure \ref{fig:description_to_TKS}.

\input{sections/figures/world_distribution}

In practice, the similarity between a vacancy and a NICE job role is computed in the full 768-dimensional sBERT embedding space using the extracted KST embeddings. The two-dimensional visualization is therefore intended only to convey the notion of relative proximity and how ranking of NICE work roles takes place. The work role embeddings depicted in Figure \ref{fig:cyberbridge-nice-comprehensive} are derived from an actual t-SNE projection of the embedding space, whereas the position of the example vacancy (marked with an $x$) is illustrative and does not represent its exact geometric location in the original embedding space. This distinction is important since distances observed in the two-dimensional visualization do not faithfully preserve the true distances from the full high-dimensional space. 

An initial observation that can be drawn from Figure \ref{fig:cyberbridge-nice-comprehensive} is that there is a clear clustering pattern within the embedding space. Work roles belonging to the same NICE workforce category tend to occupy nearby regions. Oversight \& Governance roles appear predominantly at the extremes, Implementation \& Operation roles are positioned more centrally, Design \& Development roles cluster on the right-hand side, and Protection \& Defense together with Investigation roles are located on the left. This spatial separation suggests that CyberBridge captures meaningful structural relationships within the NICE framework, as roles within the same category are represented closer to each other than to roles outside their category. 

In Figure \ref{fig:cyberbridge-nice-comprehensive}, all NICE work roles are color-coded by work role category. The three highest-ranked roles for the example vacancy as computed by CyberBridge are Incident Response, Systems Security Analysis, and Executive Cybersecurity Leadership. Since CyberBridge is \textit{interpretable} and white-box by design, it is possible to trace back which KST descriptions contributed most strongly to these assignments. For Incident Response, the vacancy Task Description `Support Cyber Security incidents as a tier 3 support' most closely matched the NICE Task Description `Perform cyber defense incident triage'. For Executive Cybersecurity Leadership, the vacancy Knowledge Description `Knowledge of security by design principles' played the largest role, aligning with `Knowledge of network security principles and practices' in the corresponding NICE role. Similarly, for Systems Security Analysis, the vacancy Task Description `Document security architecture' most closely matched the NICE Task Description `Document systems security activities'. This illustrates how competency-level semantic alignment shapes the work role ranking. 

In contrast, explanations produced by DeepSeek-V3 largely restate the high-level summaries of the corresponding NICE work roles rather than pointing to specific competency-level alignments between the vacancy text and the NICE framework. For example, DeepSeek-V3 assigned Digital Forensics as the most related work role, and when prompted for an explanation, produced the following: `Direct match to `Perform forensic investigation in a case of an incident' and `Hands-on experience with computer forensics'. Role involves analyzing digital evidence from security incidents.' While this explanation may appear plausible, it is important to recognize that explanations generated by LLMs after a prediction are typically \textit{post-hoc} rationalizations. They provide a justification for the output but do not necessarily reflect the internal reasoning process that led to the prediction. As such, the generated explanation cannot be taken as evidence that the classification was actually motivated by the factors stated in the explanation.

Furthermore, a closer inspection of the explanation reveals that the reasoning primarily reiterates sentence segments from the vacancy description that include the word `forensics', rather than explaining why the competencies required for the vacancy align with the competencies associated with the Digital Forensics role in the NICE framework. Consequently, the explanation largely paraphrases information already present in the input information, making it only superficially explanatory instead of providing meaningful reasoning about competency-level alignment. This limitation is expected, as DeepSeek only has access to the short summary statement of the role, being `Responsible for analyzing digital evidence from computer security incidents to derive useful information in support of system and network vulnerability mitigation.', lacking detailed knowledge of what the work role entails in terms of KST.

 \subsection{Job demand per region} \label{sec:job_demand}

Next, we examine how CyberBridge can be used to identify regional differences in cybersecurity workforce demand. As a proof of concept, CyberBridge was applied the combined REWIRE and LinkedIn dataset. 
The geographic distribution of this dataset is depicted in Figure \ref{fig:world}. 

To analyze regional demand, we estimated the prevalence of each NICE work role category at the continental level for North America and Europe. Other regions were excluded due to insufficient sample sizes, as the dataset contained only 55 vacancies from Australia and a single vacancy from Afghanistan. The resulting distribution of work role categories is shown in the bottom-left (North America) and middle-left (Europe) panels of Figure \ref{fig:world}. The results suggest that Protection and Defense and Investigation roles are more prevalent in Europe, whereas Implementation and Operation roles appear more prominent in the North American job market. This illustrates how CyberBridge can be used to explore regional differences in workforce demand, thereby addressing RQ2.  

Several considerations should be taken into account when interpreting these results. The dataset used in this analysis constrains the extent to which conclusions about current workforce demands can be drawn, both because of its modest size ($\approx 7,000$ vacancies) and because most postings were collected between 2022 and 2024. Consequently, these results should be interpreted primarily as a proof of concept demonstrating the methodological feasibility of CyberBridge for regional workforce analysis, rather than as an accurate representation of the present-day cybersecurity labor market. 



\subsection{Case study: Curriculum planning with CyberBridge} \label{sec:case_study}
\input{sections/figures/case_study}

We now address RQ3 by illustrating how CyberBridge can support curriculum planning by linking curricula to labor market demand. To demonstrate this, we consider the following case study.
\textit{After completing the mandatory components of her MSc degree in Cybersecurity at Nanyang Technological University, Eve is preparing for a career in Europe. Which electives should she take in her final semester to best position herself for the European cybersecurity job market? }

The overall workflow is illustrated in Figure \ref{fig:case_study}. We begin by examining the core component of the NTU curriculum,  which consists of the courses `Computer Security', `Application Security', `Cryptography', and `Security \& Risks Management'. The goal of this case study is to identify a set of electives such that the resulting curriculum aligns as closely as possible with the requirements of the European cybersecurity job market.
Following CurricuLLM \cite{nijdam2026CLLM}, we adopt Knowledge Areas as a common frame of reference between the curriculum and the job market. KAs represent cybersecurity competency domains relevant to university education, as defined in the CSEC2017 framework \cite{csec2017}. For example, the mandatory NTU courses contain a relatively large contribution of Data Security, while placing less emphasis on Human and Societal Security, which is consistent with the technical focus suggested by the titles. 

Next, we estimate the KA distribution required by the European cybersecurity labor market. To this end, all 1,105 European vacancies from the combined REWIRE and LinkedIn dataset are processed using CyberBridge (top left of the figure). For each vacancy, CyberBridge selects the most relevant NICE work role, which are subsequently aggregated to estimate the overall demand for NICE roles in the European cybersecurity job market (top middle). 
We then employ Table E.15 from CurricuLLM \cite{nijdam2026CLLM} to translate NICE roles into a desired Knowledge Area distribution.
By weighting the KA distributions associated with each NICE role according to the demand estimated by CyberBridge, we obtain an aggregated KA profile reflecting current labor market demand. This profile is visualized as a pie chart on the right side of Figure \ref{fig:case_study}, with the KA color coding shown at the bottom left. 

When comparing the required KA profile with the KA distribution of the core NTU curriculum, it appears that the core curriculum places relatively strong emphasis on Data Security, while Connection Security is comparatively underrepresented. To address this gap, we perform an exhaustive search over all electives listed in the official NTU curriculum \cite{ntu_msc_cybersecurity} to identify the combination of courses that, together with the core curriculum, most closely matches the required KA profile. 

The resulting curriculum (depicted on the bottom right) demonstrates improved alignment with the European workforce demand as expressed through the required KA profile. In particular, the selected electives `Network Security', `Security Monitoring \& Threat Detection', `Topics in Crypto and Cybersecurity', and `Blockchain \& Cryptocurrency' increase coverage of Connection Security while reducing the dominance of Data Security. However, the Miscellaneous KA remains underrepresented due to limited course availability, suggesting that this competency area could be addressed through electives offered outside the standard degree programme.

By using CyberBridge, Eve's curriculum has become more closely aligned with the competency profile demanded by the European cybersecurity market. This case study illustrates how CyberBridge can be combined with CurricuLLM to translate market demands into actionable curriculum recommendations.  

%% file: sections/figures/world_distribution.tex
\begin{figure*}[htb]
    \centering
    \resizebox{0.8\textwidth}{!}{
    \begin{tikzpicture}
    
    \begin{scope}[scale=0.8]
        \tikzset{set state val/.style args={#1/#2}{#1={fill=brown!#2}}}
        \tikzset{set state val/.list={
            UnitedStatesOfAmerica/100, Canada/70,  UnitedKingdom/70, CzechRepublic/70, Lithuania/50,France/50,
            Germany/40, Spain/50, Italy/40, Belgium/50, Portugal/50, 
            Greece/50, Australia/50, 
            Latvia/30, Netherlands/30, Poland/30, Serbia/30, 
            Bulgaria/10, Croatia/10, Estonia/10, Finland/10, Ireland/10, Luxembourg/10, Norway/10, Romania/10, Slovakia/10, Slovenia/10, Switzerland/10, Afghanistan/10
        }}
        
        \WORLD[every state={draw=white, thin, fill=black!20}]
        \coordinate (EuropeSW) at (Portugal.south west);
        \coordinate (EuropeNE) at (Finland.north east);
        
        \draw[red, thick, dashed] 
            (EuropeSW) ++(-0.5,-0.5) rectangle 
            (EuropeNE) ++(1.0,0.8);
        \node[red, font=\small\bfseries] at 
            ($(EuropeNE)!0.5!(EuropeSW)+(0.5,0.8)$) {Europe};

        \coordinate (NASW) at (UnitedStatesOfAmerica.south west);
        \coordinate (NANE) at (Canada.north east);
        
        \draw[blue, thick, dashed] 
            (NASW) ++(0,-0.5) rectangle 
            (NANE) ++(1.5,1.0);
        \node[blue, font=\small\bfseries] at 
            ($(NANE)!0.5!(NASW)+(0.2,1.2)$) {North America};
        
    \end{scope}

    \begin{scope}[scale=0.8,shift={(21,-6)}] 
    \draw (0.15,0) rectangle (3.7,3.1);
    \node[anchor=north west, font=\small] at (0.2,3.05) {\textbf{Sample Density}};
    
    \fill[brown,opacity=1.0] (0.4,2.45) rectangle (0.9,2.15);
    \node[anchor=west, font=\small] at (1.1,2.25) {$>$ 1,000};
    
    \fill[brown,opacity=0.7] (0.4,2.0) rectangle (0.9,1.7);
    \node[anchor=west, font=\small] at (1.1,1.80) {100--999};
    
    \fill[brown,opacity=0.5] (0.4,1.55) rectangle (0.9,1.25);
    \node[anchor=west, font=\small] at (1.1,1.35) {50--99};
    
    \fill[brown,opacity=0.4] (0.4,1.1) rectangle (0.9,0.8);
    \node[anchor=west, font=\small] at (1.1,0.9) {10--49};
    
    \fill[brown,opacity=0.2] (0.4,0.65) rectangle (0.9,0.35);
    \node[anchor=west, font=\small] at (1.1,0.45) {1--9};
    
    \end{scope}

\begin{scope}[scale=0.8, shift={($(Algeria)!0.5!(Sudan)+(-1,-4.65)$)}]
        
        
        \filldraw[fill=white, draw=black, thick] (0,-1) rectangle (6.3,3);
        \node[font=\small\bfseries] at (3,2.5) {European Job Categories};
        
        \fill[blue!60] (0.4,0.4) rectangle (1.2,1.698); 
        \fill[green!60] (1.6,0.4) rectangle (2.4,1.055); 
        \fill[orange!60] (2.8,0.4) rectangle (3.6,0.568); 
        \fill[purple!60] (4.0,0.4) rectangle (4.8,0.698); 
        \fill[red!60] (5.2,0.4) rectangle (6.0,0.483); 
        
        \node[font=\small, anchor=north west] at (0.35,0.2) {OG};
        \node[font=\small, anchor=north west] at (1.55,0.2) {DD};
        \node[font=\small, anchor=north west] at (2.75,0.2) {IO};
        \node[font=\small, anchor=north west] at (3.95,0.2) {PD};
        \node[font=\small, anchor=north west] at (5.15,0.2) {IN};
        
        \node[font=\tiny] at (0.8,1.848) {51.9\%};
        \node[font=\tiny] at (2.0,1.205) {26.2\%};
        \node[font=\tiny] at (3.2,0.718) {6.7\%};
        \node[font=\tiny] at (4.4,0.848) {11.9\%};
        \node[font=\tiny] at (5.6,0.633) {3.3\%};

        \node[font=\small] at (3,-0.65) {Total: 1,105 positions};
        
    \end{scope}

    \begin{scope}[scale=0.8, shift={($(UnitedStatesOfAmerica)+(-2,-6.2)$)}]
        
        
        \filldraw[fill=white, draw=black, thick] (0,-1) rectangle (6.3,3);
        \node[font=\small\bfseries] at (3,2.5) {North American Job Categories};
        
        \fill[blue!60] (0.4,0.4) rectangle (1.2,1.738);
        \fill[green!60] (1.6,0.4) rectangle (2.4,1.06);
        \fill[orange!60] (2.8,0.4) rectangle (3.6,0.665);
        \fill[purple!60] (4.0,0.4) rectangle (4.8,0.615);
        \fill[red!60] (5.2,0.4) rectangle (6.0,0.423);
        
        \node[font=\small, anchor=north west] at (0.35,0.2) {OG};
        \node[font=\small, anchor=north west] at (1.55,0.2) {DD};
        \node[font=\small, anchor=north west] at (2.75,0.2) {IO};
        \node[font=\small, anchor=north west] at (3.95,0.2) {PD};
        \node[font=\small, anchor=north west] at (5.15,0.2) {IN};
        
        \node[font=\tiny] at (0.8,1.888) {53.5\%};
        \node[font=\tiny] at (2.0,1.21) {26.4\%};
        \node[font=\tiny] at (3.2,0.815) {10.6\%};
        \node[font=\tiny] at (4.4,0.765) {8.6\%};
        \node[font=\tiny] at (5.6,0.573) {0.9\%};

        \node[font=\small] at (3,-0.65) {Total: 5,781 positions};
        
    \end{scope}
    
    \end{tikzpicture}
    }
    \caption{Sample Density of combined REWIRE and LinkedIn datasets, containing approximately 7,000 cybersecurity vacancies. The legend for the sample density is displayed on the right. The overall distribution of work role categories for North America and Europe are depicted in the bottom of the figure. }
    \label{fig:world}
\end{figure*}
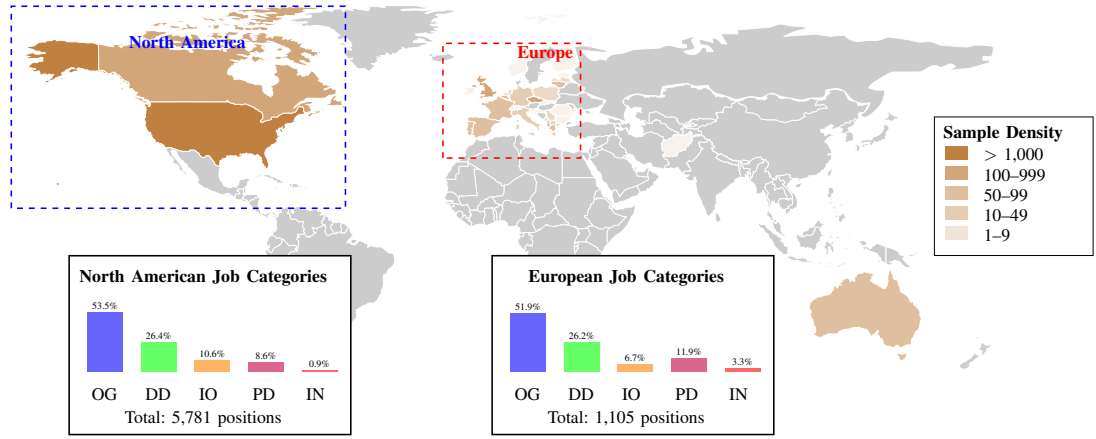

%% file: sections/figures/case_study.tex
\newcommand{\pieradius}{1.45}

\begin{figure*}[htb]
    \centering
    \resizebox{0.75\textwidth}{!}{%
    \begin{tikzpicture}[descbox/.style={draw, text width=0.2\textwidth, align=justify, font=\scriptsize, inner sep=3mm,fill=white},
    jobbox/.style={draw, rounded corners=3pt, minimum height=8mm, minimum width=4.5cm, align=center, font=\scriptsize},
    sbertblock/.style={
        rectangle,
        draw=gray!70!black,
        thick,
        fill=gray!20,
        rounded corners=3mm,
        text width=2cm,
        align=center,
        minimum width=2cm,
        minimum height=1.5cm,
        font=\sffamily\bfseries,
        inner sep=5mm
    },
    ]



\node[descbox] (jobdesc) {
\textbf{Senior Cybersecurity Engineer}\\
Reporting to the Cyber Security Team Lead, the Senior Cybersecurity Engineer will primarily focus on forensics, log management and threat modeling. The Cybersecurity Engineer is a seasoned professional, proactive self-starter who has strong problem-solving analytical skills with great attention to detail. The candidate must be comfortable working in a fast-paced, entrepreneurial, goal-oriented environment, emphasizes accountability for delivering results, and has hands-on experience with the latest security processes.
Responsibilities
Collaborate with senior management across the company to prioritize security initiatives.
Establish control-related processes/procedures while working towards building relevant security metrics and dashboards

};
\node[descbox, below right=-30mm and 5mm of jobdesc.west] (jobdesc2) {
\textbf{Chief Information Security Officer
}\\
TP Global Operation Ltd (trading under the name Truphone www.truphone.com) is a UK based but globally active, specialized telecommunications services provider that has MVNO licenses in 9 countries and operations in 15 countries. It is a world leader in providing recorded mobile communication services to regulated financial institutions and counts the world’s largest and most prestigious financial institutions like Goldman Sachs, JP Morgan and HSBC as its clients. In addition, Truphone provides core mobile communications services to a number of multinational businesses like Netflix and Tesla and IoT connectivity solutions to over 20 million devices. 
};
\node[descbox, below right=-30mm and 5mm of jobdesc2.west] (jobdesc3) {
\textbf{Cyber Security Consultant}\\
A career within Cybersecurity and Privacy services will provide you with the opportunity to help our clients implement an effective cybersecurity programme that protects against threats, propels transformation, and drives growth. As companies pivot toward a digital business model, exponentially more data is generated and shared among organisations, partners and customers. We play an integral role in helping our clients ensure they are protected by developing transformation strategies focused on security, efficiently integrate and manage new or existing technology systems to deliver continuous operational improvements and increase their cybersecurity investment, and detect, respond, and remediate threats. 

};

\node[above=0.05cm of jobdesc] {European Cybersecurity Vacancies};

\node[jobbox, fill=blue!40,below right=-73mm and 40mm of jobdesc3] (job31) {\textbf{Security Control Assessment = 31\%}};

\node[jobbox, fill=orange!40, below=1mm of job31] (job11) {\textbf{Cybersecurity Architecture = 9\%}};

\node[jobbox, fill=blue!40, below=1mm of job11] (job37) {\textbf{Systems Security Management = 8\%}};


\node[jobbox, fill=green!40, below=2mm of job37] (job28) {\textbf{Secure Systems Development = 7\%}};
\node[jobbox, fill=red!40, below=2mm of job28] (job29) {\textbf{Defensive Cybersecurity = 5\%}};
\node[below=0mm of job29] (dots1) {$\vdots$};
\node[jobbox, fill=blue!40, below=10mm of job29] (job27) {\textbf{Technology Program Auditing = 0\%}};
\node[above=0.05cm of job31] {European NICE Role distribution};


\node[sbertblock, right=1cm of jobdesc2, anchor=west] (bridge) {CyberBridge};
\node[sbertblock, right=9.5cm of jobdesc2, anchor=west] (cllm) {CurricuLLM};

\node[right= 1.75cm of cllm] (REQUIRED) {};
\pie[
    pos=REQUIRED.center,
    radius=\pieradius,
    color={KA6,KA9,KA1,KA2,KA3,KA4,KA5,KA7,KA8},
    sum=auto
]{
18/,
9/,
6/, 
3/, 
12/, 
6/, 
6/, 
32/, 
8/ 
}


\node[align=left,font=\footnotesize,below=2.5cm of jobdesc] (kabox) {
      \textbf{KA Legend:}\\
   $\color{KA6}\bullet$ Miscellaneous\\
   $\color{KA9}\bullet$ Data\\
    $\color{KA1}\bullet$ Software\\
    $\color{KA2}\bullet$ Component\\
    $\color{KA3}\bullet$ Connection\\
    $\color{KA4}\bullet$ System\\
    $\color{KA5}\bullet$ Human\\
    $\color{KA7}\bullet$ Organizational \\
    $\color{KA8}\bullet$ Societal
};

\draw[thick, gray!70, rounded corners=5pt] ([xshift=-5pt, yshift=5pt]kabox.north west) rectangle ([xshift=5pt, yshift=-5pt]kabox.south east);

\node[above=1.5cm of REQUIRED] {Required KA profile};

\node[below=5.5 cm of bridge] (NTU_CORE) {};
\pie[
    pos=NTU_CORE.center,
    radius=\pieradius,
    color={KA6,KA9,KA1,KA3,KA4,KA5,KA7,KA8},
    sum=auto
]{
4/,
28/, 
10/, 
5/, 
9/, 
14/, 
26/, 
4/
}

\node[above=1.5cm of NTU_CORE] {Completed NTU courses};

\node[sbertblock, right=2.5cm of NTU_CORE, anchor=west] (llm) {Choose 4 electives};

\node[right=2.5 cm of llm] (COMBINED) {};
\pie[
    pos=COMBINED.center,
    radius=\pieradius,
    color={KA6,KA9,KA1,KA2,KA3,KA4,KA5,KA7,KA8},
    sum=auto
]{
4/,
15/, 
9/, 
2/, 
16/, 
9/, 
10/, 
30/, 
4/
}
\node[above=1.5cm of COMBINED] {Total curriculum};

\draw[thick, gray!70, rounded corners=5pt] ([xshift=-60pt, yshift=60pt]NTU_CORE.north west) rectangle ([xshift=75pt, yshift=-100pt]COMBINED.south east);
\draw[thick, dotted, gray!50!black, -{Latex[length=2mm, width=1.5mm]}] ([yshift=-1.4cm]REQUIRED.south) -- ++(0,-2cm) -| (llm.north);

\draw[thick, dotted, gray!50!black, -{Latex[length=2mm, width=1.5mm]}] ([xshift=1.4cm]NTU_CORE.east) -- ++(0,0) |- (llm.west);

\draw[thick, dotted, gray!50!black, -{Latex[length=2mm, width=1.5mm]}] (llm.east) -- ++(0,0) |- ([xshift=-1.6cm]COMBINED.west);

\node[below=1.7cm of COMBINED, xshift=0cm, yshift=-0cm, align=left, font=\footnotesize] (selected) {
\textbf{Selected Courses: } \\
$\bullet$ Network Security \\
$\bullet$ Security Monitoring \& Threat Detection  \\
$\bullet$ Topics in Crypto \& Cybersecurity \\
$\bullet$ Blockchains \& Cryptocurrency 
};

\node[below=1.7cm of NTU_CORE, xshift=0cm, yshift=-0cm, align=left, font=\footnotesize] (selected) {
\textbf{Mandatory Courses: } \\
$\bullet$ Computer Security \\
$\bullet$ Application Security  \\
$\bullet$ Cryptography \\
$\bullet$ Security \& Risks Management 
};

    \end{tikzpicture}

}
    \caption{Procedure taken for the Case Example. All European cybersecurities included in our dataset are passed through CyberBridge to compute the prevalence of each NICE work role. Then, through CurricuLLM, the required distribution of Knowledge Areas for a European job is computed (top right). In the bottom of the figure, we illustrate how Eve can sample her curriculum at NTU such that the total curriculum matches the required KA profile as closely as possible.}
    \label{fig:case_study}
\end{figure*}
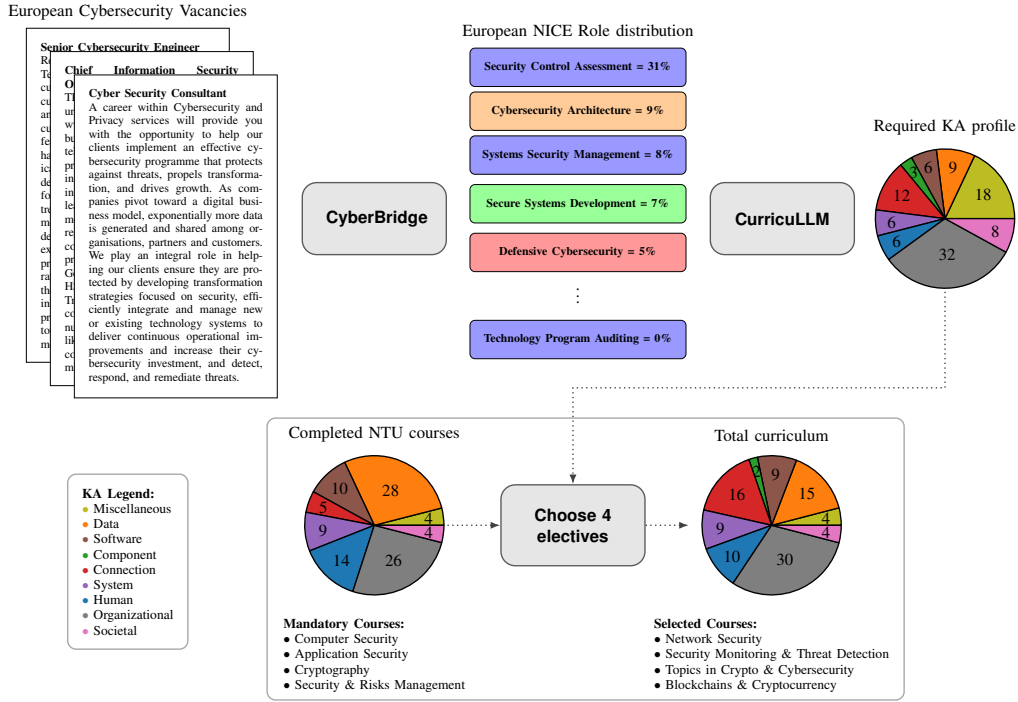

%% file: sections/discussion.tex
\section{Discussion} \label{sec:discussion}

Limitations and opportunities for future work are: 


\textbf{Data Collection}. Pairing CyberBridge with current workforce data represents an important direction for future work. In particular, the methodology described in Section \ref{sec:job_demand} could be applied larger and more recent datasets. One approach would be to automatically collect job advertisements that are currently online by scraping platforms such as LinkedIn, or official government job portals e.g., USAJobs. Continuous data collection would make it possible to analyze temporal trends in cybersecurity workforce demand. For instance, this would enable monitoring how demand for specific NICE work roles evolves over time. 
An additional validation opportunity would be to attempt to replicate the statistics reported by CyberSeek \cite{cyberseek}, which provides detailed U.S. demand estimates per NICE work role and category. 

\textbf{Human Evaluation}. Further human evaluation would strengthen the empirical validation of CyberBridge. Most importantly, these expert annotations could be used to assess CyberBridge's capabilities in mapping job advertisements to NICE work roles with greater confidence. This would also enable empirically determining optimal values for the weights $w^K, w^S\text{, and } w^K$, as well as the similarity threshold $\tau$, which could lead to a more accurate and robust pairing.  

\textbf{Job Recommendation}. While the current methodology provides the foundation for CyberBridge as a Job Recommendation System, realizing this potential requires implementing CyberBridge as a tool and conducting user evaluation. In particular, the approach described in Section \ref{sec:case_study} could allow students to filter cybersecurity vacancies in real time based on NICE role preferences or competencies from their curriculum. This would enable evaluation of user experience and CyberBridge’s effectiveness in supporting career planning. 

%% file: sections/conclusion.tex
\section{Conclusion} \label{sec:conclusion}

The paper introduced CyberBridge, an ontology-based framework that connects cybersecurity job advertisements, workforce frameworks, and academic curricula. By leveraging the Knowledge, Skill and Task descriptions of the NICE framework together with the semantic similarity-based sBERT model, CyberBridge enables interpretable mappings between vacancies and standardized cybersecurity work roles (RQ1). 

We further demonstrated how CyberBridge can support labor market analysis by identifying differences in the prevalence of NICE work role categories across geographic regions (RQ2). While these results should be interpreted as a proof of concept, they illustrate how the framework can be used to explore trends in the cybersecurity job market.

Finally, by integrating CyberBridge with the CurricuLLM framework, we showed how labor market insights can be linked to academic curricula (RQ3), supporting curriculum planning and career guidance for cybersecurity students. Overall, CyberBridge provides a practical foundation for bridging the gap between cybersecurity education and workforce needs.

